# An evaluation of Bradfordizing effects

Philipp Mayr[1]

28 August 2008

## Abstract

The purpose of this paper is to apply and evaluate the bibliometric method Bradfordizing for information retrieval (IR) experiments. Bradfordizing is used for generating core document sets for subject-specific questions and to reorder result sets from distributed searches. The method will be applied and tested in a controlled scenario of scientific literature databases from social and political sciences, economics, psychology and medical science (SOLIS, SoLit, USB Köln Opac, CSA Sociological Abstracts, World Affairs Online, Psyndex and Medline) and 164 standardized topics. An evaluation of the method and its effects is carried out in two laboratory-based information retrieval experiments (CLEF and KoMoHe) using a controlled document corpus and human relevance assessments. The results show that Bradfordizing is a very robust method for re-ranking the main document types (journal articles and monographs) in today's digital libraries (DL). The IR tests show that relevance distributions after re-ranking improve at a significant level if articles in the core are compared with articles in the succeeding zones. The items in the core are significantly more often assessed as relevant, than items in zone 2 (z2) or zone 3 (z3). The improvements between the zones are statistically significant based on the Wilcoxon signed-rank test and the paired T-Test.

## 1  Introduction

The background for the research is that distributed search across multiple databases will automatically generate large and heterogeneous document sets for subject-specific questions (Tenopir, 1982, Hood and Wilson, 2001). As a result, users have to deal with a huge amount of documents from different scientific domains, as well as for specific research topics. The perceived expectations of users searching the web are that system architects should list the most relevant or important documents in the result list first. More and more approaches appear that draw on advanced methods to produce qualitative results and alternative views on document spaces. Google PageRank and Google Scholar's citation count are just two popular examples for informetric-based mechanisms applied in Internet search engines. Similar techniques can and should be applied in DL to satisfy user demands. This paper should be seen as an argument and example for alternative re-ranking methods applied in text-based retrieval systems.

In 2004, the German Federal Ministry for Education and Research funded a major terminology mapping initiative at the GESIS Social Science Information Centre in Bonn (GESIS-IZ) "Competence Center Modeling and Treatment of Semantic Heterogeneity" (KoMoHe), which concluded in 2007 (see Mayr and Petras, 2008). The task of the KoMoHe project was to organise, create and manage "cross-concordances" between major controlled vocabularies and to evaluate DL models.

---
[1] GESIS Social Science Information Centre (GESIS-IZ), Lennéstr. 30, Bonn, Germany.
philipp.mayr@gesis.org

The KoMoHe project was the starting point and background of the following approach. The paper focuses on the evaluation of experiments with Bradfordizing (White, 1981) as a re-ranking method in DL.

An extensive review of the literature of Bradford Law of Scattering (BLS) is provided by Lockett (1989) and Wilson (1995). BLS is a well known bibliometric law which has received a lot of attention in information and library science research (e.g. Vickery, 1948; Brookes, 1968, 1969; Garfield, 1971; Buckland, 1972; Bonitz, 1980). BLS describes how articles in a subject (topic) are scattered across journals. "*If scientific journals are arranged in order of decreasing productivity of articles on a given subject, they may be divided into a nucleus of periodicals more particularly devoted to the subject and several groups or zones containing the same number of articles as the nucleus, when the numbers of periodicals in the nucleus and succeeding zones will be as a : n : n$^2$ : n$^3$ ...*"" (Bradford 1948). There are numerous of examples of applications of Bradford law in various disciplines like natural sciences and social sciences (e.g. Wagner-Döbler, 1997; Peritz, 1990). This law seems to be a very robust and commonly appearing phenomenon in most of the current literature databases and bibliographies. BLS is still under discussion as shown by recent papers (Nicolaisen and Hjørland, 2007, Mayr and Umstätter, 2007, Umstätter, 2005, Bates, 2002, Hood and Wilson, 2001).

The paper by Bates (2002) is interesting in our context because it brings together BLS and information seeking behaviour or IR. "*... the key point is that the distribution tells us that information is neither randomly scattered, nor handily concentrated in a single location. Instead, information scatters in a characteristic pattern, a pattern that should have obvious implications for how that information can most successfully and efficiently be sought.*" (Bates, 2002) Bates applies conceptually different search techniques (directed searching, browsing and linking) to the classical three Bradford zones. She postulates the utilization of the Bradford nucleus (core) for browsing, the following zone (z2) for directed searching with search terms and further zones (z3) for linking.

The intent in our approach is an automatic change between directed searching (searching with controlled terms enhanced by treatment of semantic heterogeneity) into browsing. Starting with a subject-specific descriptor search (see step 1 Identify in Figure 1), we will treat the query with our heterogeneity modules (Mayr and Petras, 2008 to appear) to transfer descriptor terms into a multi-database scenario. In a second step, the result lists from the different databases are combined and sorted according to Bradford's method (most productive journals for a topic first). After this step we have a bradfordized list of journal articles (see Table 1 for an example). Step 3 is the extraction of a result set of all documents in the Bradford nucleus which can be delivered for browsing. This browsing modus with is based on automatically bradfordized lists can be compared to Bates search technique "journal run".

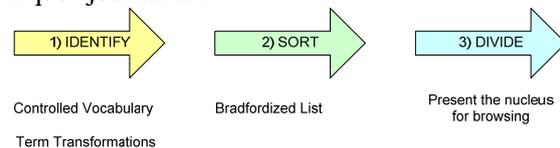

Figure 1: Operationalization of Bradfordizing

## 2   Research questions

In the next chapters we try to answer the following research questions:

1) Is a re-ranking of documents according to the Bradford law (journal productivity) an added value for users? The re-ranking of content to the most frequent sources (extracting the nucleus) can for example be a helpful access mechanism for browsing (Bates, 2002) and initial search stages. The evaluation of the utility of such a mechanism is still a desideratum.

2) Are the documents in the nucleus of a bradfordized list (core journals show a high productivity for a topic) more relevant for a topic than items in succeeding zones with a lower productivity? A study by Pontigo and Lancaster (1986) concluded that less productive journals are not necessarily of lower quality but mostly less cited. This has to be proven on a larger scale by intellectual assessments (analogous to the TREC or CLEF studies) of different user groups (e.g.

experts, novice searchers, information scientists).

3) Can Bradfordizing be applied to document sources other than journal articles? A paper by Worthen (1975) and our own analyses show that monograph literature can be successfully bradfordized. But is this a utility? Other document types (proceedings, grey literature etc.) have to be equally proven.

Table 1: Example of core journals (bradfordized list) in the field of informetrics (extracted from the LISA database, see Mayr and Umstätter, 2007)

| Journal | No. of papers |
|---|---|
| Scientometrics | 1,413 |
| Journal of the American Society for Information Science | 218 |
| Nauchno Tekhnicheskaya Informatsiya | 110 |
| Revista Espanola de Documentacion Cientifica | 96 |
| Journal of Information Science | 87 |
| Information Processing and Management | 79 |
| Journal of Documentation | 75 |
| Annals of Library Science and Documentation | 66 |

4) Can Bradfordizing be used to create an alternative view on search results? Compared to traditional text-oriented ranking mechanisms, our informetric re-ranking method offers a completely new view on results sets (see e.g. Table 1), which have not been implemented and tested in heterogeneous database scenarios with multiple collections to date.

## 3 Methods

We focus on a mix of methodologies:

1. Bradfordizing as a sorting mechanism for databases in our distributed scenario. White explains the procedure: "... *That is sorting hits (1) by the journal in which they appear, and then sorting these journals not alphabetically by title but (2) numerically, high to low, by number of hits each journal contains. In effect, this two-step sorting ranks the search output in the classic Bradford manner, so that the most productive, in terms of its yield of hits, is placed first; the second-most productive journal is second; and so on, down through the last rank of journals yielding only one hit apiece.*" (1981: p. 47). In our study we analyzed scientific literature from social and political sciences, economics, psychology and medical science databases (SOLIS, SoLit, USB Köln Opac, CSA Sociological Abstracts, World Affairs Online, Psyndex and Medline) in exactly this way. 1) Searching documents, 2) sorting or re-ranking document via Bradfordizing, 3) Dividing documents into three equally-sized zones (compare Figure 1).

2. Intellectual assessments of document relevance have been performed following the classical IR evaluation experiments at TREC (Harman and Voorhees, 2006) and Cross-Language Evaluation Forum (CLEF[2]) (Petras et al., 2007). That followed an empirical analysis of the results for subject-specific topics and questions. We retrieved, analyzed and assessed 164 different standardized topics which result in more than 96,000 documents from all above domains (see Table 2 and appendix with a typical topic and a document, listing 1, 2). More then 51,000 assessed documents could be bradfordized.

3. The utility of the nucleus/core has been investigated also in a simple user test.

## 4 Preliminary results

The preliminary results present parts of the results. In the following (result 1, 3 and 4) we will concentrate on one sample (25 topics) from the domain-specific track at CLEF 2005. The other samples in CLEF and KoMoHe show very similar results.

**Result 1**: Bradford distributions appear in all subject domains and also for results of scientific literature databases. It follows that Bradfordiz-ing can be used for re-sorting results, generally for topic-specific queries.

---

[2]http://www.gesis.org/en/research/information_technology/CLEF_DELOS.htm

Table 2: Summary of the analysed topics and documents in the information retrieval experiment (CLEF and KoMoHe)

|  | CLEF | KoMoHe |
|---|---|---|
| Period | 2003-2007 | 2007 |
| Topics | 125 | 39 |
| Documents total | 65,297 | 31,155 |
| Documents bradfordized | 29,157 | 22,332 |
| Journal article | 18,112 | 17,432 |
| Monographs | 11,045 | 4,900 |
| Databases involved | 2 | 7 |

Figure 2 and 3 show the typical scattering of documents across documents. All analyzed document collections and document types (journal articles and monographs) in our experiments show similar distributions.

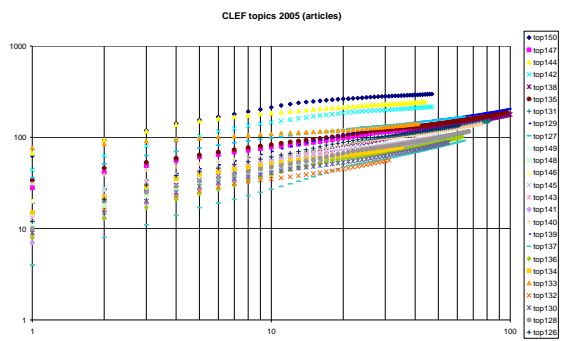

Figure 2: Bradford graphs (log-log plot) for 25 topics from the CLEF evaluation 2005. Journals are on the x-axis, journal articles are cumulated on the y-axis. Documents mean: 142, Journals total mean: 61, core mean: 4.64, z2 mean: 17.24, z3 mean: 39.56.

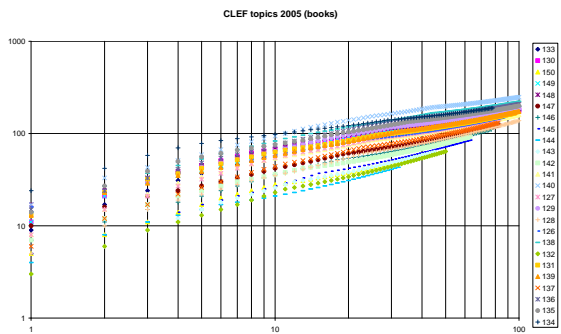

Figure 3: Bradford graphs (log-log plot) for the same 25 topics from the 2005 sample. Publishers are on the x-axis, monographs are cumulated on the y-axis. Documents mean: 211, Publishers total mean: 90, core mean: 8.60, z2 mean: 29.56, z3 mean: 51.88.

In Figure 2 each zone (core, zone 2 = z2 and zone 3 = z3) consists of approximately 47 articles. The documents are scattered over 61 journals: the highest concentration is in the core with ~5 journals, z2 consists of ~17 journals and the 47 articles in z3 are scattered across ~40 journals). In Figure 3 each zone (core, z2 and z3) consists of approximately 70 monographs. The documents are scattered over 90 publishers: the highest concentration is in the core with ~9 publishers, z2 consists of ~30 publishers and the 70 monographs in z3 are scattered across ~52 publishers).

**Result 2**: The application of informetric methods for re-ranking of documents can produce an alternative view of a result set. Intuitively non-expert users rated this view/re-ordering as positive (compare White, 1981). Positive is generally the novelty and insight which comes up when presenting highly cited papers, papers of central authors (Mutschke, 2003), articles from core journals (see Table 1) and the relevance distribution of the newly organized result set. Our interviews with experts and non-experts (12 persons) in 24 social sciences topics show clearly that the presentation of core journals after Bradfordizing is a value-added for both types of users.

**Result 3**: The application of Bradfordizing or the core journal re-ranking for subject-specific document sets leads to significant improvements of the precision between the three Bradford zones. The core journals cover significantly more relevant documents than journals in zone 2 or zone 3. The largest increase in precision can typically be observed between core and zone 3 (see Figure 4).

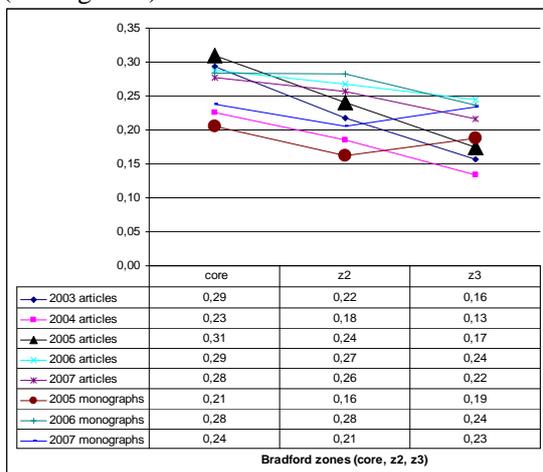

| | core | z2 | z3 |
|---|---|---|---|
| 2003 articles | 0,29 | 0,22 | 0,16 |
| 2004 articles | 0,23 | 0,18 | 0,13 |
| 2005 articles | 0,31 | 0,24 | 0,17 |
| 2006 articles | 0,29 | 0,27 | 0,24 |
| 2007 articles | 0,28 | 0,26 | 0,22 |
| 2005 monographs | 0,21 | 0,16 | 0,19 |
| 2006 monographs | 0,28 | 0,28 | 0,24 |
| 2007 monographs | 0,24 | 0,21 | 0,23 |

**Bradford zones (core, z2, z3)**

Figure 4: Precision values for items (articles and monographs) in core, z2 and z3 (125 topics from the CLEF evaluation, compare Table 2). The both data sets CLEF 2005 journal articles and monographs (documents in Figure 2 and 3) are displayed with larger data points.

**Result 4**: Bradfordizing for concentrated result sets can be successfully applied for monographs (publisher as sorting criterium, see Figure 3). The application of Bradfordizing or the core publisher re-ranking for monographs leads in general to lower improvements (compared with journal re-ranking) of the precision distribution between the three zones (core, z2 and z3, see Figure 4 and Table 3).

**Result 5**: The results show that articles in core journals are valued more often relevant then articles in succeeding zones. This stands in opposition to the original conception of relevance distribution in the zones by Bradford. This result can probably be explained with a) core journals publish more state-of-the-art articles, b) core journals are more often peer-reviewed and c) core journals cover more aspects of the searched topic than journals in the peripheral zones.

**Result 6**: The results show that the journals in the core appear approximately monthly while journals in the succeeding zones appear bi-monthly.

Table 3: Baseline, z3 and improved precision for articles and monographs in the core. Mean values for 25 topics from the CLEF 2005 dataset. The improvements between the zones core and z3 (articles) and core and baseline are statistically significant (*) based on the Wilcoxon signed-rank test and the paired T-Test. Improvements between core and z3 and core and baseline monographs are positive but not statistical significant.

| | Precision | Improvement |
|---|---|---|
| all articles (baseline) | 0.239 | |
| articles in core | 0.310 | (29.52%)* |

| | Precision | Improvement |
|---|---|---|
| articles in z3 | 0.174 | |
| articles in core | 0.310 | (78.03%)* |

| | Precision | Improvement |
|---|---|---|
| all monographs (baseline) | 0.188 | |
| monographs in core | 0.205 | (8.98%) |

| | Precision | Improvement |
|---|---|---|
| monographs in z3 | 0.188 | |
| monographs in core | 0.205 | (9.09%) |

Table 3 shows precision improvements (mean values for 25 topics) between different document clusters (baseline and core and additionally z3 and core). Baseline means all documents in the sample. The mean precision of all articles (baseline) is 0.239 whereas precision in the core is 0.310 and z3 is 0.174. According to this the core is improving baseline (29.52%) and

z3 (78.03%). Lower improvement can be observed in the monograph section.

## 5    Conclusion

Bradfordizing can successfully be applied in a set of scientific literature databases. Bradfordizing holds true in different domains and document types. A value-added for this re-ranking method can be empirically demonstrated in terms of precision improvements on a significant level. Users are intuitively satisfied with the re-ranked results. The results can also be seen as a concretion of Bradford Law in so far as Bradford did not postulate or observe a relevance advantage in the core. We can empirically show such an advantage when applying and evaluating Bradfordizing in a controlled scenario.

Further research will focus on the implementation and evaluation of the method in a live system with different modules to improve retrieval (see Mayr et al, 2008 as an example). A next step will be the exploration of the side effects and bias (see for example Nicolaisen and Hjørland, 2007) of this promising re-ranking method. The application of other evaluation methods (e.g. evaluation of full texts instead of metadata) would be highly desired.

## 6    Appendix

```
<top>
<num>163</num>
<EN-title>Risk behavior</EN-title>
<EN-desc>Research papers and publications on
risk behavior among children and adoles-
cents</EN-desc>
<EN-narr>Types of risk behavior (drug use, smok-
ing, alcohol, violence, tests of courage,       vio-
lations of the law). How are these types explained,
what measures are recommended,       and what
is the future prognosis? What preventive measures
are evaluated?</EN-narr>
</top>
```

Listing 1: A typical tagged topic description.

```
<DOC>
<DOCID>iz-solis-90128016</DOCID>
<IDENTIFIER1>19900100914</IDENTIFIER1>
<ISSN>0172-6404</ISSN>
<TITLE-DE>Historical research in the age of the com-
puter</TITLE-DE>
<DOCTYPE>journalarticle</DOCTYPE>
<SOURCE>Historical social research Quantum-
Information</SOURCE>
```

```
<MEDIATYPE/>
<AUTHOR>Schurer, K.</AUTHOR>
<PUBLICATION-YEAR>1985</PUBLICATION-
YEAR>
<LANGUAGE-CODE>en</LANGUAGE-CODE>
<CONTROLLED-TERM-
DE>Anwendung#EDV#historische Sozialfor-
schung#Informationstechnologie#Instrumentarium</CO
NTROLLED-TERM-DE>
<ABSTRACT-DE>'Computers are a useful research
tool that historians have only recently acquired. The
advantages of speed and consistancy that computers can
offer to analytical study are well known. Yet to what
degree is there a potential danger of research becoming
hindered by a misuse of technology? If the computer
using historian is to avoid problems of inflexibility he
should not allow research to be straight - jacketed by
either the computer or its software. Lastly, historians
should be aware of possible consequences that the pre-
sent revolution in information technology may have on
future research.' (author's abstract)</ABSTRACT-DE>
<TEXT/>
</DOC>
```

Listing 2: A typical tagged document in the advanced GIRT4 format.

## Acknowledgement

The project "Competence Center Modeling and Treatment of Semantic Heterogeneity" at GESIS-IZ was funded by BMBF, grant no. 01C5953. See project website for more information.
http://www.gesis.org/en/research/information_technology/komohe.htm

I would like to thank my colleague Vivien Petras who pointed me at the assessed topics from CLEF evaluation 2003-2007 and our assisting student Dirk Hohmeister who helped with the assessments and analysis.